\documentclass{article}

\usepackage{arxiv}
\usepackage{xcolor}
\usepackage[utf8]{inputenc}
\usepackage[T1]{fontenc}
\usepackage{hyperref}
\usepackage{url}
\usepackage{booktabs}
\usepackage{amsfonts}
\usepackage{amsmath}
\usepackage{nicefrac}
\usepackage{microtype}
\usepackage{graphicx}
\usepackage{natbib}
\usepackage{doi}
\usepackage{amssymb}

\title{Interaction-Conditional Semantics and the Dissolution of Quantum Paradoxes}

\author{
  Jonathon Sendall\\
  OU Philosophy Department\\
  \texttt{jonathon.sendall@ou.ac.uk}\\[0.3em]
  \href{https://jonathonsendall162367.substack.com/p/the-painted-ball-and-the-rules-of}{\textcolor{blue}{Substack Article}}
}

\begin{document}
\maketitle

\begin{abstract}
This paper argues that several canonical puzzles in quantum mechanics, including spin measurement, the double slit, entanglement correlations, and Wigner's friend, share a common origin in a semantic error: the illicit promotion of interaction-conditional outcomes to intrinsic properties. I introduce four principles that license only configuration-relative predication, grounding outcomes in physical measurement geometry while preserving objectivity. Applying these principles uniformly dissolves each puzzle without new physics or ad hoc interpretive machinery. Bell's theorem and the Kochen-Specker theorem are reframed not as dynamical mysteries but as constraints on permissible explanatory structure, evidence that intrinsic-outcome semantics is incompatible with empirical reality. The result is a relational objectivity that avoids both naive property realism and observer-subjectivism.
\end{abstract}

\section{Introduction: The Semantic Error in Quantum Measurement}

Quantum mechanics is remarkable for the precision of its predictions and the persistence of its paradoxes. Nearly a century after the theory's formalisation, physicists and philosophers continue to debate whether particles have definite properties before measurement, whether measurement causes physical collapse, whether entangled systems communicate instantaneously, and whether observers play a constitutive role in determining outcomes. These debates are often framed as problems requiring new physics, new mathematics, or new ontology. This paper argues that they require none of these. The puzzles share a common origin in a semantic error: the illicit promotion of interaction-conditional outcomes to intrinsic properties.

The error is easy to make and difficult to see. When a Stern-Gerlach apparatus registers ``spin-up,'' the natural report is ``the particle has spin-up.'' When a photon strikes a detector behind a double slit, we say the photon ``went through'' one slit or the other. When entangled particles yield correlated results at spacelike separation, we ask how one particle ``knows'' what the other did. In each case, the language treats an outcome, something produced by a specific interaction between system and apparatus, as though it were a property the system possessed independently of that interaction.

This is not a minor imprecision. It is a category error with systematic consequences. Once an interaction outcome is mistaken for an intrinsic property, contradiction follows. If the particle has spin-up, then a measurement along a different axis that yields spin-down seems to require the particle to have changed, or to have possessed incompatible properties simultaneously. If the photon went through one slit, then the interference patterns would become inexplicable. If entangled particles carry correlated values, then Bell's theorem forces a choice between hidden variables and nonlocal influence. The puzzles multiply because the same mis-typed predication recurs across different experimental contexts.

To clarify the scope of this proposal, it is helpful to state explicitly what interaction-conditional semantics is and what it is not.

\begin{itemize}
    \item It is not a proposal for observer-dependent reality. The ``viewpoints'' described are physical configurations, not mental states.
    \item It is not a modification of the quantum formalism. It introduces no new equations, no new dynamical collapse parameters, and no new empirical predictions that deviate from standard quantum mechanics.
    \item It is not a denial of Bell's theorem or a redefinition of locality. It accepts the theorem's conclusions fully but diagnoses the violation differently than hidden-variable theories do.
    \item It is a diagnosis of a systematic type error in standard quantum discourse: the application of outcome-predicates to systems in isolation rather than to system-configuration pairs.
    \item It is a demonstration that spin, slit, entanglement, and Wigner puzzles share a common semantic origin and dissolve under a unified logical discipline.
\end{itemize}

The proposal developed here is that these puzzles dissolve once the predication error is corrected. The correction is not a new interpretation of quantum mechanics in the sense of a rival ontological picture. It is a disciplined constraint on permissible statements, a semantic hygiene that specifies when outcome-predicates may be legitimately applied. The core rule is simple: outcome predicates are licensed only as interaction-conditional predicates. ``The particle yields spin-up under this measurement orientation'' is well-formed. ``The particle has spin-up'' is not false; it is ill-typed. It attempts to assign a relational outcome as though it were a monadic property.

This rule is not eliminative about outcomes. Outcomes are real, objective, and repeatable. What the rule denies is that they are intrinsic to the measured system in isolation. They are intrinsic to the system-apparatus configuration. This is relational objectivity: the outcome is objective because any observer who instantiates the same physical configuration will obtain the same result; it is relational because the configuration is a necessary part of the specification.

A crucial clarification distinguishes this view from observer-relative or subjectivist interpretations. ``Viewpoint,'' in the sense used here, does not mean perspective, belief, or consciousness. It means measurement configuration: the orientation of a detector, the geometry of an experimental arrangement, the physical boundary conditions under which an interaction occurs. Viewpoints are not interchangeable descriptions of the same fact; they are mutually exclusive physical setups. From one configuration, the outcome licensed by another configuration does not exist, not as hidden information, not as a fact awaiting discovery, but as an interaction that has not occurred.

The philosophical payoff of this semantic discipline is a unified diagnosis across multiple canonical puzzles. Spin measurement, the double slit, entanglement correlations, and Wigner's friend all exhibit the same structural pattern: an interaction-conditional outcome is treated as intrinsic, and apparent paradox results \citep{Wigner1961}. In each case, restoring the conditionality dissolves the paradox without introducing new dynamical mechanisms or ad hoc interpretive machinery. The formalism of quantum mechanics is left intact; what changes is how we describe what the formalism is about.

This paper proceeds as follows. Section 2 introduces a simple model, the painted ball, not as a physical mechanism but as a heuristic for correct predication. Section 3 states the proposal as explicit principles, defining the ontology of structure and configuration and offering a formal typing discipline. Section 4 applies these principles to spin measurement, the double slit, entanglement, and Wigner's friend. Section 5 addresses Bell's theorem and the Kochen-Specker theorem, framing them not as dynamical mysteries but as constraints on admissible predication \citep{Bell1964, KochenSpecker1967}. Section 6 situates the view relative to Bohr, relational quantum mechanics, and contextuality approaches. Section 7 responds to objections. Section 8 concludes with implications for practice.

\section{The Painted Ball: A Predication Model}

Before stating the principles formally, it helps to train intuition with a simple model.

\paragraph{Disclaimer:} The model presented here is a heuristic for interaction-conditional predication only. It is a classical analogy intended to train intuition regarding the distinction between fixed structure and geometric outcomes. It does not model non-commutativity, superposition, or entanglement, nor does it represent a physical mechanism for quantum states. Its sole purpose is to illustrate valid predication logic.

Imagine a solid ball floating in space. Its surface is coated with an unusual paint, one that reflects different colours depending on the angle from which it is viewed. The coating is real and fixed; it does not change while anyone looks at it. Yet, standing in one position, you see the ball as blue, while someone else, standing elsewhere, sees it as red.

Both observations are correct and both are repeatable. If each observer stays in place, they will keep seeing the same colour. Swap positions, and they swap what they see. Crucially, the ball itself has not changed at all.

The mistake would be to say ``the ball is blue'' or ``the ball is red.'' Such statements treat an interaction outcome as though it were an intrinsic property. The correct description is always conditional: the ball appears blue from here; it appears red from there. The colour is not stored in the ball as a single fact. It is produced by the interaction between the fixed surface, the incoming light, and the physical position of the observer. Once we recognise this, an apparent contradiction, two people reporting different colours of the same object, dissolves into straightforward geometry.

There is a subtlety worth pausing on. From position A, the view from position B is not merely unknown; it is undefined relative to that configuration. Person A cannot access what Person B sees by thought, calculation, or reinterpretation. To see what Person B sees, Person A must physically move through space and time and adopt the geometry of position B. The blue view does not secretly contain the red view, nor is the red view missing information that would reveal the blue. Each is a different physical interaction that cannot occur without a change in geometry. Viewpoints are not interchangeable descriptions; they are mutually exclusive physical configurations.

To speak clearly about such situations, it helps to think of the world as operating under a system of rules that separates what is fixed, the ball and its surface structure, from what is produced by interaction, the colour seen from a given location. This system does not decide outcomes, does not choose, and does not depend on belief or awareness. It simply specifies which interactions are allowed and what outcomes are admissible under each. The ball never violates these rules, and neither do the observers.

\section{Principles of Interaction-Conditional Semantics}

The proposal can be stated as a set of explicit principles. Each principle is a constraint on how outcome-predicates may be used in descriptions of quantum systems.

\subsection{The No-Intrinsic-Outcome Principle}
Outcome predicates are never attributed to a system simpliciter. They are attributed only to system-apparatus configurations. ``The particle has spin-up'' is not false but ill-formed; it attempts to predicate of the system alone what can only be predicated of the system under a specified measurement orientation. The licensed statement is: ``The particle yields spin-up under measurement orientation $\theta$.''

\subsection{The Physical-Configuration Principle}
``Viewpoint'' means instrument orientation, coupling geometry, and boundary conditions, not psychology, consciousness, or observer belief. A configuration is a macroscopically stable measurement configuration (apparatus plus environment) that fixes an interaction basis or pointer regime and is repeatably re-instantiable. Configurations are physical facts about the experimental arrangement; they can be specified without reference to any observer's mental state. This principle blocks the slide from relationality to subjectivism.

\subsection{The Relational Objectivity Principle}
Outcomes are objective if and only if they are invariant under repetition of the same configuration across observers, locations, and times. Objectivity does not require intrinsicality. A property can be fully objective, public, repeatable, law-governed, while being essentially relational.

\subsection{The State-as-Bookkeeping Principle}
Wavefunctions and quantum states are permitted as compact encodings of objective constraints on admissible outcome-probability maps, not as material fields requiring physical collapse. ``Collapse'' redescribes a change in the admissible interaction geometry, not a dynamical event in which a physical entity discontinuously transforms. The state encodes the rules for how systems respond to configurations, without mandating that the system possesses the response prior to the configuration.

\subsection{The Structure-Configuration Ontology}
To ground these semantic constraints, we must specify the ontology they describe. If outcomes are not intrinsic properties, what exists?

First, Configuration ($C$). A configuration is not a subjective perspective. It is a macroscopically stable physical arrangement, an apparatus coupled to a relevant environment, that establishes a fixed basis for interaction. It corresponds to the physical regime in which decoherence has effectively diagonalised the density matrix, defining a stable set of pointer states. The requirement that configurations be macroscopically stable is not arbitrary. It reflects the empirical fact that measurement outcomes correlate with pointer states that have decohered relative to environmental degrees of freedom \citep{Zurek2003}. Decoherence provides the physical mechanism by which definite outcome-bases emerge, though it does not by itself determine which outcomes obtain.

Second, Structure. The quantum state represents objective structure, but this structure is relational. It is not a hidden variable that encodes pre-existing outcome values. Rather, it is a constraint pattern that determines which outcome probabilities are admissible for any given system-configuration pair. We need not decide here whether this structure is dispositional, nomological, or structural-realist in nature. The minimal commitment required is that the structure is real, objective, and resides in the relation between the system and its potential configurations, not in the system alone.

\subsection{Formal Précis: A Typing Discipline for Outcomes}
The proposal can be summarised formally as a strict typing discipline for outcome predicates.

\begin{description}
    \item[Rule 1 (Admissibility):] An outcome statement is admissible only if it is typed as $O(S,C)$ with $C$ explicitly specified. The form $O(S)$ is ill-typed and semantically inadmissible.
    \item[Rule 2 (Composition):] Any claim that combines outcomes from incompatible configurations $C_1$ and $C_2$ into a single descriptive account is inadmissible unless a bridging interaction is specified. This bridging interaction must instantiate a new configuration $C^*$ in which comparison is licensed.
\end{description}

\paragraph{Examples of Bridging Interactions:}
\begin{itemize}
    \item Wigner opens the lab door and reads the friend's notebook (creates $C^*$ where both outcomes are jointly accessible).
    \item Alice and Bob meet and compare measurement records (creates $C^*$ where correlation becomes manifest).
    \item A detector measures ``which slit?'' (creates $C^*$ incompatible with interference, preventing mixing of slit-outcomes with phase information).
\end{itemize}

\paragraph{Not legitimate bridging:}
\begin{itemize}
    \item Alice ``calculates what Bob should see'' without interaction (attempts to access $O(S, B, C, B)$ from within $C, A$).
    \item Wigner ``deduces'' the friend's result from the external quantum state (conflates structure with outcome).
\end{itemize}

The bridging interaction must be physical, not merely epistemic.

For any valid pair $(S,C)$, the quantum state $\psi$ defines an admissible set of outcomes and a probability measure $P(O \mid S,C)$ according to the Born rule. Objectivity is defined as the stability of this probability measure and the resulting frequencies under repeated instantiations of $C$.

Crucially, if two configurations $C_1$ and $C_2$ are incompatible (physically mutually exclusive), there is no joint probability distribution $P(O_1, O_2 \mid S)$ defined over them. This is not missing information; it is non-definition due to the physical mutual exclusivity of measurement configurations. Requesting such a joint distribution is not a request for unknown data; it is a category error. This refusal of joint distributions over incompatible contexts is exactly the feature required to avoid the Kochen-Specker contradictions.

\subsection{Probability and the Born Rule}
Under this view, probabilities are objective features of the system-configuration pairing. They do not represent ignorance of hidden values, nor are they merely subjective degrees of belief. They represent the objective propensity of the shared structure to yield specific outcomes under specific configurations. The Born rule is taken as given; it is part of the structural package encoded by the state. This framework organises and interprets the probability calculus; it does not attempt to derive it from non-probabilistic foundations.

One might object: couldn't probabilities represent ignorance of configuration-dependent facts that are nevertheless determinate before measurement? This would be a configuration-relative hidden variable theory. Reply: This possibility is ruled out by the same contextuality results. Even if we relativise values to configurations, we cannot assign them consistently across overlapping contexts (KS), nor can we reproduce the observed correlations through local configuration-relative values (Bell). The probabilities are irreducibly part of the structure itself, not placeholders for missing information.

\subsection{Glossary of Key Terms}
\begin{description}
    \item[Configuration ($C$):] Macroscopic, decohered experimental arrangement specifying boundary conditions for interaction; equivalent to a stable coupling regime between system and apparatus.
    \item[Structure ($\psi$):] Objective constraint pattern on admissible configuration-outcome pairs, encoded in the quantum state; not a set of pre-existing values.
    \item[Outcome ($O$):] Physical result of system-configuration interaction; only well-defined relative to a specific configuration.
    \item[Shared structure:] Non-separable constraint pattern for composite systems; grounds correlations without requiring communication.
    \item[Bridging interaction:] Physical process that instantiates a new configuration allowing previously incompatible outcomes to be jointly accessed.
    \item[Intrinsic-outcome semantics:] (Rejected) The view that systems possess outcome-values independently of configuration.
    \item[Relational-outcome semantics:] (Endorsed) The view that outcome-predicates apply only to system-configuration pairs.
\end{description}

\section{Case Studies}

The power of interaction-conditional semantics lies in its uniform applicability. The same principles, applied without modification, dissolve puzzles that are often treated as requiring distinct solutions.

\subsection{Spin Measurement}
A particle is prepared and sent through a Stern-Gerlach apparatus oriented along the $z$-axis. The apparatus registers ``spin-up.'' If the particle is instead measured along the $x$-axis, it may yield ``spin-down.'' Under interaction-conditional semantics, the particle does not have spin-up or spin-down as intrinsic properties. The particle yields spin-up under $z$-axis measurement and spin-down under $x$-axis measurement. These are not incompatible properties of the particle; they are distinct outcomes of distinct configurations.

\subsection{The Double Slit}
The double slit experiment is often presented as showing that particles are ``sometimes waves and sometimes particles.'' Under this proposal, ``wave-like'' and ``particle-like'' are not identities the particle switches between. They are outcome-types produced by different interaction geometries. Open geometry admits interference; constrained geometry (detectors at the slits) admits localisation. The detector is not an observer in any psychological sense; it is a physical constraint that alters the configuration. No decision, no awareness, no paradox.

\subsection{Entanglement}
Entanglement seems stranger still. Two particles are created together and then separated. When one is measured, the other shows a correlated result.

There are indeed two particles, and they are spatially separate. Each measurement interaction is local. But what was created is not two independent systems with separate states. The pair shares a single structure that governs how outcomes appear under interaction. Entanglement correlations are the manifestation of this shared structure.

The holistic structure constrains admissible joint outcomes relative to joint configurations; it does not introduce a propagating influence variable. It is important to note: Taken literally, the painted ball picture from Section 2 would amount to a Bell-local hidden variable model, and thus could not reproduce quantum correlations. The analogy holds only for the logic of predication (structure vs.\ outcome). In entanglement, the ``shared structure'' is not a set of hidden values carried by the particles (screening off the correlation), but a global constraint on the admissible outcomes of the joint configuration. The correlation does not live in space or travel between particles; it lives in the global structure governing the pair. To use an imperfect analogy: just as the Pythagorean theorem constrains the relationship between sides of a right triangle without any side `influencing' another, the entangled structure constrains the relationship between distant outcomes without any outcome influencing another.

\subsection{Wigner's Friend}
Wigner's friend stages a conflict between observers. Inside a sealed laboratory, a physicist (the ``friend'') measures a particle. Outside, Wigner treats the laboratory as a superposition.

The painted ball resolves the tension. The mistake is assuming one viewpoint must subsume the other. The friend's outcome is produced by one interaction geometry: friend-plus-apparatus-plus-particle. Wigner's description corresponds to a different interaction geometry: Wigner-plus-sealed-laboratory.

Crucially, from Wigner's configuration, the friend's outcome is not ``hidden'' or ``unknown''; there is no fact of the matter defined for Wigner's configuration regarding the friend's specific result. Paradoxes of the Frauchiger-Renner type arise precisely when one attempts to mix outcome predicates from these incompatible configurations into a single logical account, violating Rule 2 (Composition) \citep{FrauchigerRenner2018}. Interaction-conditional semantics strictly forbids this cross-configuration predication. Reconciliation requires a physical bridging interaction that instantiates a new configuration $C^*$ (e.g., opening the door), after which outcome predicates may be re-typed under $C^*$.

\section{Bell's Theorem and Kochen-Specker Constraints}

Bell's theorem and the Kochen-Specker (KS) theorem are often viewed as the ultimate barriers to any realist interpretation of quantum mechanics that remains local \citep{Bell1964, KochenSpecker1967}. Interaction-conditional semantics offers a different reading: these theorems are metalogical constraints that reveal the impossibility of intrinsic-outcome semantics.

\subsection{The Assumptions}
Bell's theorem relies on several assumptions to derive its inequalities:
\begin{enumerate}
    \item \textbf{Outcome Determinism (Value Definiteness):} The system possesses definite values for outcomes prior to measurement.
    \item \textbf{Locality (Factorizability):} Probabilities at one wing depend only on local settings and local variables.
    \item \textbf{Measurement Independence:} Settings are chosen independently of hidden variables.
\end{enumerate}
Standard Hidden Variable theories attempt to preserve Assumption 1 (outcomes are pre-existing properties) and are thus forced by Bell's result to abandon Assumption 2 (Locality).

\subsection{What This Regime Changes}
Interaction-conditional semantics rejects Assumption 1, not because outcomes are ``fuzzy,'' but because the attribution of an outcome value to a system in isolation is ill-typed. The statement ``the particle pair carries definite correlated values'' is formally invalid in this framework.

Consequently, there is no map $v(A), v(B)$ of values waiting to be revealed. Because there are no pre-existing local outcome values, the failure of Bell's inequalities does not imply that influence has travelled faster than light to change those values. It implies that the correlations reflect a non-separable, global structure established at the source that constrains outcome-pairs $(O_A, O_B)$ relative to configuration-pairs $(C_A, C_B)$ without any need for causal connection between the measurements.

\paragraph{What Grounds the Correlations: A Worked Example}
Consider the singlet state measured at angles $\alpha$ (Alice) and $\beta$ (Bob).
A local hidden variable theory would posit: $\lambda$ encodes $v_A(\theta)$ and $v_B(\theta)$ for all $\theta$, with correlation $= \int v_A(\alpha) \cdot v_B(\beta) \rho(\lambda) d\lambda$.

This framework instead posits: The structure $\psi_{\text{singlet}}$ directly constrains the 4-tuple space $(\alpha, \beta, O_A, O_B)$. The constraint is: $P(\uparrow_\alpha, \downarrow_\beta \mid \alpha, \beta) = \frac{1}{2} \sin^2[(\alpha-\beta)/2]$.

Key difference:
\begin{itemize}
    \item \textbf{LHV:} Values exist independently at each wing; $\lambda$ screens off the correlation.
    \item \textbf{ICS:} No independent wing-values exist; the constraint is primitive and holistic.
\end{itemize}

The correlation is not produced by the measurements nor transmitted between them. It is manifest when the joint configuration $(\alpha, \beta)$ is instantiated, because the structure was laid down as a unity at the source. This is not action-at-a-distance because there is no action, no change in a physical quantity at one wing caused by events at the other. It is non-separability: the constraint pattern cannot be factorised into independent constraints on each wing.

\subsection{The Kochen-Specker Lesson}
The Kochen-Specker theorem proves that for Hilbert spaces of dimension $\ge 3$, no non-contextual value assignment to all observables can respect quantum algebraic constraints. Specifically: if you try to assign values $v(A), v(B), v(C)$ to observables such that $v(f(A,B)) = f(v(A), v(B))$ for all functional relations, you reach contradiction.

In interaction-conditional semantics, this is exactly what we should expect. KS attempts to construct a global assignment $v: \text{Observables} \to \text{Values}$ independent of measurement context. But our framework denies that such assignments are well-typed. There is no $v(A)$; there is only $O(S, C_A)$ where $C_A$ is a configuration that measures $A$.

The contradiction arises because KS tries to maintain value-definiteness across incompatible contexts simultaneously. Our typing discipline forbids precisely this: outcomes from incompatible configurations cannot be combined (Rule 2) unless a bridging interaction is specified.

KS is thus not a limitation on our knowledge of pre-existing values, but proof that the logical form ``the system has value $v$ for observable $A$'' independent of configuration is incoherent. This provides independent support for the No-Intrinsic-Outcome Principle (3.1).

\subsection{Relational Structure, Not Nonlocality}
The correlations observed in entanglement are grounded in the objective, shared structure of the pair. This structure is holistic; it defines the admissible joint outcomes for joint configurations. It does not screen off the correlations because it is not a local hidden variable. The Bell correlations are the signature of a reality where properties are interaction-dependent, not evidence of superluminal causation. The choice is not between Locality and Realism, but between Intrinsic-Outcome Realism and Relational-Outcome Realism. This paper chooses the latter.

\section{Situating the View}

Interaction-conditional semantics shares features with several existing interpretations but is distinct in its specific semantic formulation and ontological commitments.

\begin{table}[h]
\centering
\caption{Comparison of Approaches}
\label{tab:comparison}
\begin{tabular}{p{4cm} p{3cm} p{4cm} p{3.5cm}}
\toprule
\textbf{Approach} & \textbf{Subject of \newline Outcome Predicates} & \textbf{Status of \newline Wavefunction} & \textbf{Objectivity} \\
\midrule
Many-Worlds (Everett) & System alone & Physical field, all branches real & Relative to branch \\
Objective Collapse (GRW) & System alone & Physical field that collapses & Intrinsic, collapse objective \\
Relational QM (Rovelli) & System-Observer relation & Information & Relative to observer \\
QBism & Agent's Experience & Agent's degrees of belief & Personal / Subjective \\
Contextual Realism (CSM) & System-Context Modality & Tool for probabilities & Objective System-Context properties \\
\textbf{This Proposal} & \textbf{System-Configuration Pair} & \textbf{Bookkeeping for Structure} & \textbf{Relational, Repeatable under C} \\
\bottomrule
\end{tabular}
\end{table}

This proposal is closest to the Contextual Realism of Auffèves and Grangier (CSM) \citep{AuffevesGrangier2016} and the Pragmatist Realism of Healey \citep{Healey2012}. Like CSM, it views outcomes as actualised only within a context. It differs primarily in offering a semantic diagnosis of the paradoxes as category errors. It is also compatible with the insights of Spekkens regarding contextuality: that the failure of non-contextual hidden variable models points to an epistemic or structural restriction rather than a bizarre ontology \citep{Spekkens2007}.

\section{Objections and Replies}

\subsection{``Isn't This Just Semantics?''}
The objection is that semantic proposals are mere verbal manoeuvres without substantive consequences. The reply is twofold.

First, semantic discipline has genuine explanatory consequences. It identifies a single type error that unifies the treatment of spin, slit, entanglement, and Wigner. It provides a principled prohibition of the counterfactual assignments that generate Bell and Frauchiger-Renner contradictions.

Second, the proposal is not merely linguistic. It asserts that the world is such that intrinsic outcome-attributions fail. This is a constraint on permissible explanatory form, not a verbal relabelling. The fact that Bell and KS contradictions appear precisely when the typing discipline is violated is evidence that this semantics reflects a structural feature of physical reality, not a mere notational preference.

\subsection{``Does It Evade Bell?''}
The worry is that interaction-conditional semantics is a sophisticated way of dodging Bell's theorem. The reply is that no evasion is attempted. Bell's theorem is accepted as a valid constraint. The framework simply does not instantiate the type of value-map (local hidden variables) that Bell's theorem rules out. Bell tells us that intrinsic-outcome semantics fails; this regime takes the failure seriously and abandons the semantics rather than seeking nonlocal rescues.

\subsection{``Does It Make States Unreal?''}
The concern is that treating states as ``bookkeeping'' amounts to anti-realism. The reply distinguishes two senses of ``real.'' The wavefunction is real as an encoding of structure; it is the objective constraint on possibilities. It is not real as a material substance. This is selective deflation in the service of relational objectivity.

\subsection{``What About Bohmian Mechanics?''}
Bohmian mechanics assigns intrinsic positions to particles, yet reproduces quantum predictions. Doesn't this refute the claim that intrinsic-outcome semantics fails?

Reply: Bohmian mechanics is subtler than this objection suggests. Yes, particles have definite positions at all times. But the dynamics, the velocities determined by the guidance equation, depend on the quantum potential, which depends on the wavefunction of the entire experimental configuration, including apparatus and environment.

The Bohmian position is intrinsic, but Bohmian velocity (and thus the trajectory that determines measurement outcomes) is thoroughly contextual. Different experimental setups yield different quantum potentials and thus different trajectories, even for particles prepared identically. Moreover, in Bohmian mechanics, only position is privileged as intrinsic. Spin, momentum, energy, all other observables are contextual, defined only relative to measurement configuration. This is precisely the structure interaction-conditional semantics endorses. The real lesson of Bohmian mechanics is that even a theory designed to restore intrinsic properties must make those properties contextually dependent in their dynamics.

\section{Conclusion: Payoff and Implications}

\subsection{Summary}
Under interaction-conditional semantics, measurement is not mysterious. It is the instantiation of a configuration and the production of the outcome admissible under that configuration. Paradoxes dissolve because they are revealed to be artifacts of illicitly projecting outcomes into systems. Objectivity survives as the repeatability of outcomes under fixed configurations.

\subsection{Implications for Practice}
The adoption of this framework has concrete implications for physics education and foundations:
\begin{description}
    \item[Pedagogy:] Students should be trained to avoid statements like ``the particle has spin-up.'' Enforcing the relational form ``yields spin-up under orientation $n$'' from the start prevents the formation of paradoxical intuitions.
    \item[Foundations:] Research programmes whose sole aim is to ``restore intrinsic properties'' or ``solve the measurement problem'' by adding dynamics are responding to a pseudo-problem. Semantic discipline suggests redirecting effort to understanding the physical basis of configuration and decoherence.
    \item[Quantum Information:] The view clarifies that quantum protocols exploit the structural constraints on configuration-outcome maps, not superluminal influence.
\end{description}

\subsection{Closing}
The success of interaction-conditional semantics in dissolving paradoxes suggests that quantum mechanics was never as strange as it appeared, only as precise as its mathematics demanded. The formalism already embodied the relational structure; we simply misread it through an inappropriate semantic lens inherited from classical physics.

What remains genuinely mysterious, why this mathematical structure, why the Born rule, why decoherence selects the bases it does, are genuine open problems for physics. But the pseudo-mysteries generated by intrinsic-outcome talk, the measurement problem, wave-particle duality, spooky action at a distance, and the observer's role in collapse can now be set aside. They were linguistic artefacts, not features of nature.

This is the dissolution, not solution, of the quantum paradoxes. The paradoxes dissolve because they were never well-formed questions. Interaction-conditional semantics does not answer ``How does measurement work?''; it reveals that this question, as typically posed, rests on a category error. The correct question is: ``What structure governs configuration-outcome patterns?'' That question has an answer: the quantum state. And that answer was there all along.

\end{document}